\documentclass[prl,twocolumn,aps,showpacs,tightenlines,superscriptaddress]{revtex4}
\usepackage{amssymb,amsmath,amsthm,epsfig,color,graphicx,times}

\usepackage{hyperref}
\usepackage{enumerate}

\def\ket#1{|{#1}\rangle}
\newcommand{\tr}{{\operatorname{Tr}\,}}

\begin{document}

\title{Classical nature of ordered phases: origin of spontaneous symmetry breaking}
\author{M. Cianciaruso}
\affiliation{Dipartimento di Fisica ``E. R. Caianiello'', Universit\`a degli Studi di Salerno, Via Giovanni Paolo II 132, I-84084 Fisciano (SA),
Italy}
\affiliation{INFN, Sezione di Napoli, I-80126 Napoli, Italy}

\author{L. Ferro}
\affiliation{INFN, Sezione di Napoli, I-80126 Napoli, Italy}
\affiliation{Dipartimento di Ingegneria Industriale, Universit\`a degli Studi di Salerno, Via Giovanni Paolo II 132, I-84084 Fisciano (SA), Italy}


\author{S. M. Giampaolo}
\affiliation{University of Vienna, Faculty of Physics, Boltzmanngasse 5, 1090 Vienna, Austria}
\affiliation{Dipartimento di Ingegneria Industriale, Universit\`a degli Studi di Salerno, Via Giovanni Paolo II 132, I-84084 Fisciano (SA), Italy}

\author{G. Zonzo}
\affiliation{Dipartimento di Fisica ``E. R. Caianiello'', Universit\`a degli Studi di Salerno, Via Giovanni Paolo II 132, I-84084 Fisciano (SA), Italy}

\author{F. Illuminati}
\thanks{Corresponding author: fabrizio.illuminati@gmail.com}
\affiliation{INFN, Sezione di Napoli, I-80126 Napoli, Italy}
\affiliation{Dipartimento di Ingegneria Industriale, Universit\`a degli Studi di Salerno, Via Giovanni Paolo II 132, I-84084 Fisciano (SA), Italy}
\affiliation{CNISM Unit\`a di Salerno, I-84084 Fisciano (SA), Italy}

\date{August 6, 2014}

\begin{abstract}
We investigate the nature of spontaneous symmetry breaking in complex quantum systems by conjecturing that the maximally symmetry-breaking quantum ground states are the most classical ones corresponding to an ordered phase. We make this argument quantitatively precise by showing that the ground states which realize the maximum breaking of the Hamiltonian symmetries are the only ones that: I) are always locally convertible, i.e. can be obtained from all other ground states by local operations and classical communication, while the reverse is never possible; II) minimize the monogamy inequality for bipartite entanglement; III) minimize quantum correlations, as measured by the quantum discord, for all pairs of dynamical variables and are the only ground states for which the pairwise quantum correlations vanish asymptotically with the intra-pair distance.

\end{abstract}

\pacs{03.67.Mn, 05.30.Rt, 75.10.Pq}

\maketitle

In the study of collective quantum phenomena, the understanding of the ordered phases associated to local order parameters relies on the key concept of spontaneous symmetry breaking. The latter is required to explain the existence of locally inequivalent ground states
that are not eigenstates of one or more symmetry operators for the corresponding many-body Hamiltonian~\cite{S2000}.
In recent years, knowledge of quantum phase transitions between symmetry-preserving and symmetry-breaking phases has been sharpened by the
application of methods and techniques originally developed in the field of quantum information~\cite{AFOV2008,VCM2008}. Various types of quantum phase transitions have been characterized by identifying the singular points in the derivatives of different measures of bipartite~\cite{OAFF2002,ON2002} and multipartite entanglement~\cite{GH2013}. Moreover, different ordered phases can be identified by looking at the factorization properties of different ground states~\cite{GAI} or by studying the behavior of the ground-state fidelity under local or global variations of the Hamiltonian parameters~\cite{ZP2006}.

Efforts have been devoted to the investigation of the behavior of the concurrence~\cite{OPM2006}, multipartite
entanglement~\cite{GH2013,ORO2006} and of the quantum discord~\cite{TRHA2013} for some specific states. However, on the whole, the complete understanding of the physical mechanism that selects the symmetry-breaking ground states in the thermodynamic limit remains an open problem~\cite{BratteliRobinson,ADZ2002}.
In complete analogy with the case of classical phase transitions driven by temperature, the pedagogical explanation of this phenomenon invokes the unavoidable presence of some local, however small, perturbing external field that selects one of the maximally symmetry-breaking ground states (MSBGSs) among all the elements of the quantum ground space. However, the implicit assumption hidden in this type of reasoning is that the MSBGSs are the most classical ones and thus the ones that are selected in real-world situations.

In the present work we promote this assumption to an explicit general conjecture on the nature of ordered quantum phases and the origin of spontaneous symmetry breaking, and we test it by comparing various measures of classicality and quantumness for symmetry-breaking and symmetry-preserving quantum ground states. Specifically, we show that, within the quantum ground space corresponding to ordered phases with nonvanishing local order parameters, the MSBGSs are the most classical ones in the sense that they are the only quantum ground states that satisfy the following three quantitatively precise criteria for each set of Hamiltonian parameters consistent with an ordered quantum phase:
\begin{enumerate}[I)]
\item {\em Local convertibility} -- All ground states are locally convertible into MSBGSs via local operations and classical communication (LOCC), while the reverse transformation is impossible;
\item {\em Entanglement sharing} -- The MSBGSs are the only ground states that minimize the residual tangle between a dynamical variable and the rest of the system. Stated otherwise, the MSBGSs are the only ground states that satisfy monogamy of entanglement, a quantum constraint on shared correlations with no classical counterpart, at its minimum among all other possible ground states;
\item {\em Quantum correlations} -- For all pairs of dynamical variables (e.g. spins) the MSBGSs are the only ground states that minimize pairwise quantum correlations, as measured by the quantum discord. Moreover, they are the only ground states whose pairwise quantum discord vanishes asymptotically as a function of the intra-pair distance.
\end{enumerate}
These three features imply that the mechanism of spontaneous symmetry breaking selects the most classical ground states associated to ordered phases of quantum matter.

In the following, we will derive results that are of general validity for all systems that belong to the same universality class of exactly solvable models that are standard prototypes for quantum phase transitions associated to spontaneous symmetry breaking, such as the $XY$ quantum spin model~\cite{S2000}. The one-dimensional spin-$1/2$ $XY$ Hamiltonian with ferromagnetic nearest-neighbor interactions in a transverse field with periodic boundary conditions reads:
\begin{equation}\label{eq:XYmodelhamiltonian}
H \! =\!-\!\sum_{i=1}^{N}\! \! \left[\!\left(\!\frac{1+\gamma }{2}\!\right) \! \sigma_i^x \sigma_{i+1}^x \!+\!
\left(\!\frac{1-\gamma }{2}\!\right)\!\sigma_i^y
\sigma_{i+1}^y \!+\! h \sigma_i^z\right]\! \;,
\end{equation}
where $\sigma_i^\mu$, $\mu = x, y, z$, are the Pauli spin-$1/2$ operators acting on site $i$, $\gamma$ is the
anisotropy parameter in the $xy$ plane, $h$ is the transverse magnetic field, and the periodic boundary conditions
$\sigma_{N+1}^\mu \!\equiv\! \sigma_1^\mu$ ensure the invariance under spatial translations.

For this class of models, the phase diagram can be determined exactly in great detail~\cite{LSM1961,BMD1970}. In the thermodynamic limit, for any $\gamma\!\in\!(0,1]$, a quantum phase transition occurs at the critical value $h_c = 1$ of the transverse field. For $h\! <\! h_c\!=\!1$ the system is ferromagnetically ordered and is characterized by a twofold ground-state degeneracy such that the $\mathbb{Z}_2$ parity symmetry under inversions along the spin-$z$ direction is broken by some elements of the ground space. Given the two symmetric ground states, the so-called even $|e\rangle$ and odd $|o\rangle$ states belonging to the two orthogonal subspaces associated to the two possible distinct eigenvalues of the parity operator, any symmetry-breaking linear superposition of the form
\begin{equation}\label{eq:groundstates}
|g(u,v)\rangle = u |e\rangle + v |o\rangle \;
\end{equation}
is also an admissible ground state, with the complex superposition amplitudes $u$ and $v$ constrained by the normalization condition \mbox{$|u|^2\!+\!|v|^2\!=\!1$}. Taking into account that the even and odd ground states are orthogonal, the expectation values of operators that commute with the parity operator are independent of the superposition amplitudes $u$ and $v$. On the other hand, spin operators that do not commute with the parity
may have nonvanishing expectation values on such linear combinations and hence break the symmetry of the Hamiltonian (\ref{eq:XYmodelhamiltonian}).

Consider observables $O_S$ that are arbitrary products of spin operators and anti-commute with the parity. Their expectation values in the superposition ground states
(\ref{eq:groundstates}) are of the form
\begin{equation}\label{eq:expectationvalue}
\langle g(u,v)|O_S|g(u,v)\rangle = u v^* \langle o|O_S|e\rangle + v u^* \langle e|O_S|o\rangle \; .
\end{equation}
Both $\langle o|O_S|e\rangle$ and $\langle e|O_S|o\rangle$ are real and independent of $u$ and $v$ and hence the expectation
(\ref{eq:expectationvalue}) is maximum for \mbox{$u\!=\!\pm v\!=\!1\!/\!\sqrt{2}$}~\cite{BMD1970}. These are the values of the
superposition amplitudes that realize the maximum breaking of the symmetry and identify the order parameter as well as the MSBGSs.

In order to discuss the entanglement and correlation properties of quantum ground states, we consider arbitrary bipartitions $(A|B)$
of the total system such that subsystem $A=\{i_1,\ldots,i_L\}$ is any subset made of $L$ spins, and subsystem $B$ is the remainder. Given any global ground state of the total system, the reduced density matrix $\rho_A$ ($\rho_B$) of subsystem $A$ ($B$) can be expressed in general in terms of the $n$-point correlation functions~\cite{ON2002}:
\begin{equation}
\label{eq:defreduce}
 \rho_{A}\!(u,v) \!= \! \frac{1}{2^L} \!\!\!\!\!\!\!
 \ \sum_{\mu_1,\ldots,\mu_L}\! \!\!\!\! \langle g(u,v)\!| \sigma_{i_1}^{\mu_1}\! \cdots \! \sigma_{i_L}^{\mu_L} \!
 |g(u,v)\!\rangle \sigma_{i_1}^{\mu_1}\! \cdots \! \sigma_{i_L}^{\mu_L} \, ,
\end{equation}
and analogously for $\rho_B$. All expectations in Eq.~(\ref{eq:defreduce}) are associated to spin operators that either commute or anti-commute with the parity along the spin-$z$ direction. Therefore the reduced density matrix $\rho_A$ can be expressed as the sum of a symmetric part $\rho_A^{(s)}$, i.e. the reduced density matrix obtained from $|e\rangle$ or $|o\rangle$, and a traceless matrix $\rho_A^{(a)}$ that includes all the terms that are nonvanishing only in the presence of a breaking of the symmetry:
\begin{equation}
\label{eq:defreduce1}
 \rho_{A}(u,v) = \rho_A^{(s)} + (uv^*+vu^*) \rho_A^{(a)} \; .
\end{equation}
Both $\rho_A^{(s)}$ and $\rho_A^{(a)}$ are independent of the superposition amplitudes $u$ and $v$, while the reduced density matrix depends on the choice of the ground state. This implies that the elements of the ground space are not locally equivalent. Explicit evaluation of expectations and correlations in symmetry-breaking ground states in the thermodynamic limit is challenging even when the exact solution for the symmetric elements of the ground space is available.

We will now sketch a method that allows to overcome this difficulty and whose general validity is not in principle restricted to the particular model considered. In order to obtain $\rho_A^{(s)}$ it is sufficient to transform the spin operators in fermionic ones and then apply Wick's theorem. Such algorithm cannot be applied to spin operators $O_A$, acting on subsystem $A$, that anti-commute with the parity. In order to treat this case we first introduce the symmetric operator $O_AO_{A+r}$, for which, by applying the previous procedure, we can evaluate $\langle e |O_A O_{A+r}| e\rangle$. Then, the desired expectation $\langle e |O_A| o\rangle$ can be computed by exploiting the property of asymptotic factorization of products of local operators at infinite separation~\cite{BMD1970,S2000,BratteliRobinson} that yields $\langle e |O_A| o\rangle = \sqrt{\lim\limits_{r \to \infty} \langle e |O_A O_{A+r}| e\rangle}$,
where the root's sign is fixed by imposing positivity of the density matrix $\rho_{A}(u,v)$. Having obtained the exact reduced density matrix $\rho_{A}(u,v)$ for all possible subsystems $A$ and superposition amplitudes $u$ and $v$, we are equipped to investigate the nature of quantum ground states with respect to the properties of local convertibility, entanglement sharing, and range of pairwise quantum correlations.

We begin by studying the property of local convertibility of quantum ground states in an ordered phase. In general, given two pure bipartite quantum states, $\ket{\psi_1}$ and $\ket{\psi_2}$, we say that $\ket{\psi_1}$ is locally convertible into $\ket{\psi_2}$ if $\ket{\psi_1}$ can be transformed into $\ket{\psi_2}$ by using only local quantum operations and classical communication (LOCC), and the aid of an ancillary entangled system~\cite{JP1999}.

This concept of local convertibility can be formalized in terms of the entire hierarchy of the R\'enyi entanglement entropies $S_\alpha(\rho_A) \equiv \frac{1}{1-\alpha}\log_2\left[{\tr{\rho_A^\alpha}}\right]$ of the reduced density operator of subsystem $A$, which provides a complete characterization of the entanglement spectrum and its scaling behavior in different quantum phases~\cite{GMASI2013}. The necessary and sufficient conditions for a bipartite state $\ket{\psi_1}$ to be locally convertible to another state $\ket{\psi_2}$ is that the inequality $S_\alpha(\psi_1) \geq S_\alpha(\psi_2)$ holds for all bipartitions and all $\alpha>0$~\cite{T2007}. Local convertibility has been recently applied to the characterization of topological order and the computational power of different quantum phases~\cite{HammaPRL2013}.
\begin{figure}[t]
\includegraphics[width=7.5cm]{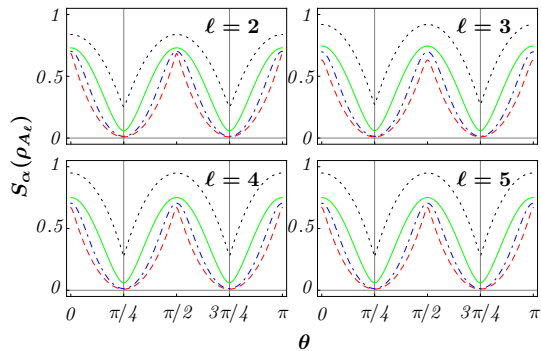}
\caption{(Color online) Behavior of the R\'enyi entropies $S_\alpha(\rho_A)$ as functions of the different ground states in the ordered phase, $h < h_c$,
in the case of a subsystem $A_{\ell}$ made of $\ell$ contiguous spins. Each line stands for a different value of $\alpha$. Black dotted
line: $\alpha=0.5$. Green solid line: $\alpha\rightarrow 1^+$ (von Neumann entropy). Blue dot-dashed line: $\alpha=3$. Red dashed line:
$\alpha\rightarrow \infty$. The different ground states are parameterized by the superposition amplitudes $u=\cos(\theta)$ and $v=\sin(\theta)$. The two vertical lines correspond to the two MSBGSs, respectively obtained for $\theta=\pi/4$ and $\theta=3 \pi/4$. The Hamiltonian parameters are set at the intermediate values $\gamma=0.5$ and $h=0.5$. Analogous behaviors are observed for different values of the anisotropy and external field.}
\label{convertibilityversusell}
\end{figure}
\begin{figure}[t]
\includegraphics[width=7.5cm]{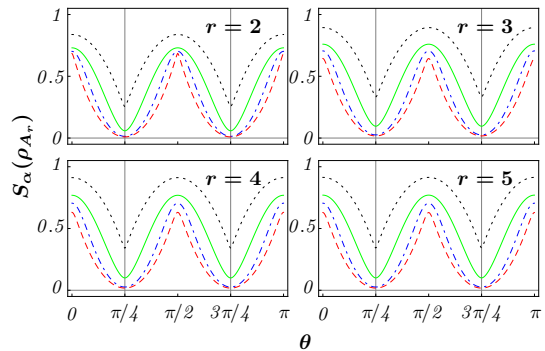}
\caption{(Color online) Behavior of the R\'enyi entropies $S_\alpha(\rho_A)$ as functions of the different ground states in the ordered phase, $h < h_c$,
in the case of a subsystem $A_{r}$ made by two spins, for different inter-spin distances $r$. Each line stands for a different value of $\alpha$. Black dotted
line: $\alpha=0.5$. Green solid line: $\alpha\rightarrow 1^+$ (von Neumann entropy). Blue dot-dashed line: $\alpha=3$. Red dashed line:
$\alpha\rightarrow \infty$. The different ground states are parameterized by the superposition amplitudes $u=\cos(\theta)$ and $v=\sin(\theta)$. The two vertical lines correspond to the two MSBGSs, respectively obtained for $\theta=\pi/4$ and $\theta=3 \pi/4$. The Hamiltonian parameters are set at the intermediate values $\gamma=0.5$ and $h=0.5$. Analogous behaviors are observed for different values of the anisotropy and external field.}
\label{convertibilityversusr}
\end{figure}

It was previously shown that symmetric ground states are always locally convertible among themselves for $h_f < h < h_c$, and never for $h < h_f < h_c$~\cite{GMASI2013}, where $h_f = \sqrt{1 - \gamma^2}$ is the factorizing field~\cite{GAI}. Here, thanks to the general methods developed in the introductory part of the present work, we are able to investigate the local convertibility property of {\em all} quantum ground states. In Fig.~\ref{convertibilityversusell} we report the behavior of the R\'enyi entropies $S_\alpha$ as functions of the different ground states for a bipartition of the system in which subsystem $A$ is made of $\ell$ contiguous spins, while in Fig.~\ref{convertibilityversusr} we report it for subsystem $A$ made of two spins with various inter-spin distances.

We observe that the MSBGSs are the ground states characterized by the smallest value of all R\'enyi entropies, independently of the size $\ell$ of the
subsystem and of the inter-spin distance $r$. Therefore, all elements in the ground space are always locally convertible to a MSBGS, while the opposite is impossible. This first quantitative criterion for classicality is thus satisfied only by MSBGSs.

We now compare symmetry-breaking and symmetry-preserving ground states with respect to entanglement sharing. The monogamy inequality quantifies in a simple and direct way the limits that are imposed on how bipartite entanglement may be distributed among many parties~\cite{CKW2000,OV2006}. For a given system of $N$ $1/2$-spins it reads:
\begin{equation}
\tau (i|N-1) \geq \sum_{j=1}^{N-1} \tau(i|j) \; \; \; \; j \neq i, \; \; \; \forall \; i \; .
\label{monogamy}
\end{equation}
In the above expression, $\tau = C^{2}$ is known as the tangle, where $C$ is the concurrence~\cite{HW1997,W1998}; the sum in the r.h.s. runs over all $N-1$ spins excluding spin $i$. The l.h.s. quantifies the bipartite entanglement between one particular, arbitrarily chosen, spin in the collection (reference spin $i$) and all the remaining $N-1$ spins. The r.h.s. is the sum of all the pairwise entanglements between the reference spin and each of the remaining $N-1$ spins. The inequality implies that entanglement cannot be freely shared among multiple quantum parties $N \geq 3$, a constraint of quantum origin with no classical counterpart.

The residual tangle $\tilde{\tau}$ is the positive semi-definite difference between the l.h.s and the r.h.s in Eq.~(\ref{monogamy}). It measures the amount of entanglement not quantifiable as elementary bipartite spin-spin entanglement. Its minimum value compatible with monogamy provides yet another quantitative criterion for classicality.

Specializing, for simplicity but without loss of generality, to translationally-invariant $XY$ spin systems in magnetically ordered phases, since the expectation value of $\sigma_i^y$ vanishes on every element of the ground space, the expressions of the tangle $\tau$ and the residual tangle $\tilde{\tau}$ for any arbitrarily chosen spin in the chain read, respectively,
\begin{eqnarray}
\tau & = & 1- m_z^2-(u^*v+v^*u)^2 m_x^2 \; , \label{tangle} \\
\tilde{\tau} & = & \tau - 2 \sum_{r=1}^{\infty} C_{r}^2(u,v) \geq 0 \label{residualtangle} \; ,
\end{eqnarray}
where $m_z\!=\!\langle e| \sigma_i^z | e \rangle\!=\!\langle o| \sigma_i^z | o \rangle$ is the on-site magnetization along $z$, the order parameter
$m_x\!=\!\langle e| \sigma_i^x | o \rangle\!=\! \sqrt{\lim\limits_{r \to \infty} \langle e| \sigma_i^x \sigma_{i+r}^x| e \rangle}$, and
$C_{r}(u,v)$ stands for the concurrence between two spins at a distance $r$ when the system is in any one of the possible ground states $|g(u,v)\rangle$,
Eq.~(\ref{eq:groundstates}).

It was previously shown that comparing the symmetric ground states with the MSBGSs, the spin-spin concurrence is larger in the MSBGSs~\cite{OPM2006} if $h < h_f < h_c$, where $h_f = \sqrt{1 - \gamma^2}$ is the factorizing field~\cite{GAI}, while for $h_f < h < h_c$ they are equal. In fact, we have verified that these two results are much more general. We have compared all ground states (symmetric, partially symmetry breaking, and MSBGSs) and we have found that for $h < h_f < h_c$ the spin-spin concurrences are maximum in the MSBGSs for all values of the inter-spin distance $r$, while for $h_f < h < h_c$ and for all values of $r$ they are independent of the superposition amplitudes $u$ and $v$ and thus acquire the same value irrespective of the chosen ground state. Finally, it is immediate to see that the third term in the r.h.s. of Eq.~(\ref{residualtangle}) is maximized by the two MSBGSs. Collecting all these results, it follows that the residual tangle is minimized by the two MSBGSs and therefore also this second quantitative criterion for classicality is satisfied only by MSBGSs.

We finally analyze the behavior of quantum correlations between any two spins for different ground states in an ordered phase. Quantum correlations are properties of quantum states more general than entanglement. Operationally, they are defined in terms of state distinguishability with respect to the so-called {\em classical-quantum} states. The latter are quantum states that, besides being separable, i.e. not entangled, remain invariant under the action of at least one nontrivial local unitary operation. In geometric terms, a {\em bona fide} measure of quantum correlations must quantify how much a quantum state {\em discords} from classical-quantum states and must be invariant under the action of all local unitary operations. A computable and operationally well defined geometric measure of quantum correlations is then the {\em discord of response}~\cite{RGI2014}. The pairwise discord of response $D_R$ for a two-spin reduced density matrix is defined as:
\begin{equation}
\label{discord}
D_R(\rho_{ij}^{(r)}(u,v)) \equiv \frac{1}{2} \min_{U_i} d_{x} \left(\rho_{ij}^{(r)}(u,v),\tilde{\rho}_{ij}^{(r)}(u,v) \right)^2 \, ,
\end{equation}
where $\rho_{ij}^{(r)}(u,v)$ is the state of two spins $i$ and $j$ at a distance $r$, obtained by taking the partial trace of
the ground state $|g(u,v)\rangle$ with respect to all other spins in the system, $\tilde{\rho}_{ij}^{(r)}(u,v)\!\equiv\! U_i\rho_{ij}^{(r)}(u,v) U_i^\dagger$ is the two-spin state transformed under the action of a local unitary operation $U_i$ acting on spin $i$, and $d_{x}$ is any well-behaved, contractive distance (e.g. Bures, trace, Hellinger) of $\rho_{ij}^{(r)}$ from the set of locally unitarily perturbed states, realized by the least-perturbing operation in the set. The trivial case of the identity is excluded by considering only unitary operations with {\em harmonic} spectrum, i.e. the fully non-degenerate spectrum on the unit circle with equispaced eigenvalues.

It has been recently shown that entropic measures of pairwise quantum discord are always nonvanishing both at finite and infinite inter-spin distance if the system is in a symmetric ground state~\cite{CRLB2013} and that they are always smaller in the MSBGSs than in the symmetric ground states~\cite{TRHA2013}. We will now show that the pairwise discord of response between any two spins is always minimized by the two MSBGSs with respect to all other possible, symmetry-preserving or symmetry-breaking, ground states. Moreover, we will show that the two MSBGSs are the only ground states for which $D_R$ vanishes asymptotically as the inter-spin distance $r \to \infty$.




In order to show that the MSBGSs are the only ground stats for which all spin pairs have vanishing quantum correlations in the limit $r \to \infty$, one needs to prove that only for such ground states there exists at least one local unitary operation $U_i$ that leaves $\rho_{ij}^{(\infty)}(u,v)$ invariant, so that
$D_R(\rho_{ij}^{(\infty)}(u,v)) = 0$. From Eq.~(\ref{eq:defreduce}) and the property of asymptotic factorization, the two-spin reduced density matrix $\rho_{ij}^{(\infty)}(u,v)$ can be conveniently written in terms of $m_i^z=\langle e|\sigma_i^z| e\rangle$ and
$m_i^x= \langle e|\sigma_i^x |o \rangle$ ($m_i^y=0$ for $\gamma > 0$). One can then easily show that the invariance relation $\tilde{\rho}_{ij}^{(\infty)}(u,v)={\rho}_{ij}^{(\infty)}(u,v)$ is verified if and only if $u=\pm v=1/\sqrt{2}$, with the corresponding unitary operation
$U_i= \cos \theta \sigma_i^z + \sin \theta \sigma_i^x$ and $\theta$ given by $\tan \theta= m_i^x/m_i^z$.
\begin{figure}
\includegraphics[width=6.5cm]{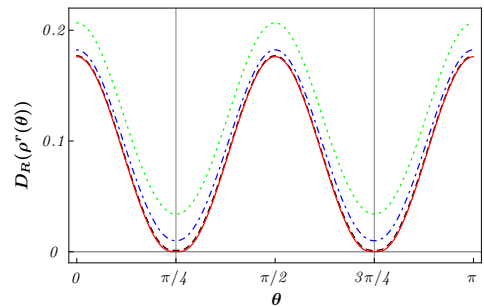}
\caption{(Color online) Behavior of the discord of response of the reduced two-spin density operator as a function of the different ground states in the ordered phase, $h < h_c$, for various distances $r$ between the two spins. Green dotted line: $r=1$. Blue dot-dashed line: $r=2$. Black dashed line: $r=3$. Red solid line: $r=\infty$. The different ground states are parameterized by the superposition amplitudes $u=\cos(\theta)$ and $v=\sin(\theta)$. The two vertical lines correspond to the two MSBGSs, respectively obtained for $\theta=\pi/4$ and $\theta=3 \pi/4$. The Hamiltonian parameters are set at the intermediate values $\gamma=0.5$ and $h=0.5$. Analogous behaviors are observed for different values of the anisotropy and external field.}
\label{quantumness}
\end{figure}
This proof holds for two-spin quantum correlations. However it can be generalized to pairwise quantum correlations with subsystem $A$ made of one spin, and subsystem $B$ made of $n$ spins that are all sufficiently far apart from each other for the property of asymptotic factorization to apply. Also in this case it is possible to prove that the quantum correlations between one spin and the remaining $n$ spins vanish if and only if the ground state of the system is a MSBGSs.

Extending the analysis to include the case of a finite inter-spin distance $r$, in Fig.~\ref{quantumness} we report the behavior of the discord of response as a function of the different possible ground states in the ordered phase for different values of $r$. From
Fig.~\ref{quantumness} we observe that among all possible ground states the MSBGSs are the ones that minimize quantum correlations for all values of $r$ and have vanishing discord for $r \to \infty$, while for all other ground state the pairwise quantum correlations remain finite for all values of the inter-spin distance, including $r \to \infty$. Therefore, the two MSBGSs associated with the maximally ordered phases are the most classical, among all the ground states of the system, in the sense that their two-spin reduced density matrices are the least discordant ones from the set of classically correlated states. Therefore also the third quantitative criterion of classicality is satisfied only by maximal spontaneous symmetry breaking.

Summarizing, we have investigated the classical nature of quantum ground states associated to ordered phases and spontaneous symmetry breaking. We have introduced three independent quantitative criteria of classicality based on local convertibility between pure quantum states, monogamy of entanglement and entanglement sharing, and the minimization and spatial range of spin-spin quantum discord. We have found that maximally symmetry-breaking ground states (MSBGSs) are the most classical among all possible ground states according to these three criteria.

These findings lend a strong quantitative support to the intuitive idea that the physical mechanism which selects the MSBGSs among all possible ground states is due to the unavoidable presence of environmental perturbations, such as local fields, which in real-world experiments necessarily drive the system onto the most classical among the possible ground states. This reasoning is strengthened by the fact that local perturbations may be described by LOCC and for each set of parameters consistent with an ordered phase all ground states are always locally convertible into the MSBGSs.

{\em Acknowledgments} - The authors acknowledge financial support from the Italian
Ministry of Scientific and Technological Research under the PRIN 2010/2011 Research Fund, and from the EU FP7 STREP Projects iQIT, Grant Agreement No. 270843, and EQuaM, Grant Agreement No. 323714. SMG acknowledges financial support from the Austrian Science Foundation, Grant FWF-P23627-N16.

\end{document}